# Searching beyond the usual papillomavirus suspects in squamous carcinomas of the vulva, penis and head and neck

Marta Félez-Sánchez[1,2], Marleny Vergara[1], Silvia de Sanjosé[1,2,3], Xavier Castellsagué[1,2], Laia Alemany[1,2] and Ignacio G. Bravo[1,2,4*]

And VVAPO/RIS HPV TT study groups

[1]Infections and Cancer Laboratory, Catalan Institute of Oncology (ICO) L'Hospitalet de Llobregat, Barcelona, Spain; [2]Bellvitge Institute of Biomedical Research (IDIBELL), L´Hospitalet de Llobregat, Barcelona, Spain; [3]Centro de Investigación Biomédica en Red: Epidemiología y Salud Pública (CIBERESP). Instituto de Salud Carlos III, Madrid, Spain; [4]MIVEGEC, National Center for Scientific Research (CNRS), Montpellier, France.

*Corresponding author contact information:
Ignacio G. Bravo
Maladies Infectieuses et Vecteurs: Ecologie, Génétique, Evolution et Contrôle (MIVEGEC)
UMR CNRS 5290, IRD 224, UM
National Center for Scientific Research (CNRS)
911 Avenue Agropolis, BP 64501
34394 Montpellier Cedex 5
France
Tlf: +33 467 41 5123
E-mail: ignacio.bravo@ird.fr

**Abstract**

Human Papillomaviruses (HPVs) are involved in the etiology of anogenital and head and neck cancers. The HPV DNA prevalence greatly differs by anatomical site. Indeed, the high rates of viral DNA prevalence in anal and cervical carcinomas contrast with the lower fraction of cancer cases attributable to HPVs in other anatomical sites, chiefly the vulva, the penis and head and neck. Here we analyzed 2635 Formalin Fixed Paraffin Embedded surgical samples that had previously tested negative for the presence of HPVs DNA using the SPF10/DEIA procedure, in order to identify the presence of other PVs not explicitly targeted by standard molecular epidemiologic approaches. All samples were reanalyzed by using five broad-PV PCR primer sets (CP1/2, FAP6064/FAP64, SKF/SKR, MY9/MY11, MFI/MFII) targeting the main PV main clades. In head and neck carcinoma samples (n=1141), we recovered DNA from two BetaHPVs, namely HPV20 and HPV21, and from three cutaneous AlphaPVs, namely HPV2, HPV57 and HPV61. In vulvar squamous cell carcinoma samples (n=902), we found one of the samples containing DNA of one cutaneous HPV, namely HPV2, and 29 samples contained DNA from essentially mucosal HPVs. In penile squamous cell carcinoma samples (n=592), we retrieved the DNA of HPV16 in 16 samples. Our results show first that the SPF10/DEIA is very sensitive, as we recovered only 2.1% (55/2635) false negative results; second, that although the DNA of cutaneous HPVs can be detected in cancer samples, their relative contribution remains anyway minor (0.23%; 6/2635) and may be neglected for screening and vaccination purposes; and third, their contribution to malignancy is not necessarily warranted and need to be elucidated.

**INTRODUCTION**

Human Papillomaviruses (HPVs) are involved in the etiology of anogenital and oropharyngeal cancers (Forman et al., 2012). Within *Papillomaviridae*, HPVs belong into five different genera, namely Alpha-, Beta-, Gamma-, Mu- and NuPVs (Bernard et al., 2010). The large majority of HPVs, essentially Beta- and GammaPVs, cause asymptomatic infections and can be detected in healthy skin swabs and for some GammaPVs, also in mucosal rinses (Bottalico et al., 2011; Gottschling et al., 2009). AlphaPVs is a very heterogeneous clade regarding tropism and clinical manifestation of the disease. Although most infections by human AlphaPVs are clinically asymptomatic some of them cause productive cutaneous warts; other cause productive mucocutaneous warts; finally a number of human AlphaPVs with mucosal tropism can induce malignant transformation after decades of persistent infection and are identified as carcinogenic or possibly/probably carcinogenic for humans (Doorbar et al., 2012; IARC, 2007).

Careful retrospective investigations have shown that infections by human AlphaPVs are the most likely etiologic cancer agent, accounting for nearly 100% of cervical cancer cases (de Sanjose et al., 2010), 88% of anal cancer cases in both males and females (Alemany et al., 2015) and for 74% of cancers of the vagina (Alemany et al., 2014). These high rates of viral DNA prevalence contrast with the lower fraction of cancer cases attributable to HPVs in other anogenital sites, chiefly the vulva and the penis. In vulvar cancers, infections by HPVs have been associated with less than 30% of the cancer cases (de Sanjose et al., 2013) while HPV DNA is found in around 30% of penile cancer (Alemany et al., 2016). Finally, in head and neck (HN) cancers, the most consistent findings relate to oropharyngeal cancers, where HPV DNA has been detected in 25% of cancer cases in contrast with the rest of the oral cavity, where HPV DNA is found in less than 10% of the cases (Castellsague et al., 2016; D'Souza et al., 2007).

The development of PCR methods using general primers for amplification of a broad-range of HPVs had a major impact on the molecular epidemiology of viral-related infections. Many different primer sets targeting the *L1* gene have been designed and used for the detection of a broad-range of HPVs, as this gene is the most conserved one at the nucleotide level (Mengual-Chulia et al., 2016). Given their overwhelming contribution to cancer, most of these consensus primers were designed to target oncogenic AlphaPVs. Among these are the single pair of consensus primers GP5/6 (Snijders et al., 1990) and its extended version GP5+/6+ (de Roda Husman et al., 1995); and the MY09/11 (Manos et al., 1989) and its extended version PGMY09/11 pair of degenerate primers (Gravitt et al., 2000). PCR methods using PGMY09/11 primers have been extensively used in epidemiologic studies of HPVs (Giuliano et al., 2001; Richardson et al., 2005; Richardson et al., 2003; Schiffman et al., 2005; Tabrizi et al., 2005). The MY-based method generates an amplicon of 450bp and targets a wide spectrum of HPVs, including all known possibly/probably/oncogenic AlphaPVs. The GP5+/6+ primer set has also been extensively used in many epidemiologic HPVs studies, either directly (Frisch et al., 1997; Hampl et al., 2006; Madsen et al., 2008; Skapa et al., 2007) or nested, after the MY primer amplification (Fox et al., 2005; Hampl et al., 2007; Piketty et al., 2003). This GP-based PCR generates an amplicon of 150bp and can amplify at least 20 mucosal AlphaPVs (de Roda Husman et al., 1995). The SPF10 primer set has been extensively used in epidemiological studies due to its high sensitivity and specificity. This method is able to amplify of 69 known AlphaPVs generating a small fragment (65bp) of the *L1* gene (Kleter et al., 1998). Tests that rely on shorter fragments of the viral genome are considered to be more sensitive and usable for less preserved specimens. None of these widely used methods of HPVs detection are generally able to detect Beta- or GammaPVs. Another system widely used is the FAP primer set, which is very useful in identifying new PVs (Forslund et al., 2002; Forslund et al., 1999). This system is able to detect a broad-spectrum of PVs from both human and animal species (Antonsson et al., 2000; Antonsson and Hansson, 2002) and is usually the choice for detecting the presence of unknown, largely divergent PVs (Antonsson and McMillan, 2006; Bzhalava et al., 2014; Garcia-Perez et al., 2014)

The aim of the present study was to reanalyze samples from squamous cell carcinomas of the HN, penis and vulva that had previously tested negative for the presence of HPVs DNA using the SPF10/DEIA procedure in order to identify the presence of other PVs not targeted by standard epidemiologic approaches, covering mucosal as well as cutaneous HPVs.

**MATERIALS AND METHODS**

*Sample Collection*

Samples were obtained from a Formalin Fixed Paraffin Embedded (FFPE) repository from a retrospective cross-sectional study coordinated by the Catalan Institute of Oncology (ICO), Barcelona, Spain, designed and constructed for the assessment of the HPV contribution to a number of anogenital and HN human tumors (Alemany et al., 2016; Alemany et al., 2015; Alemany et al., 2014; Castellsague et al., 2016; de Sanjose et al., 2013; de Sanjose et al., 2010). All specimens were tested for the presence of tumour tissue as well as for the presence of HPV DNA using a two-step SPF10/DEIA/LiPA25 protocol (Kleter et al., 1999). Amplification products testing positive for the presence of HPVs DNA but that resulted negative for LiPA25 genotyping were Sanger-sequenced to identify the nature of the viral DNA amplified. The detailed protocols and results for cervical (de Sanjose et al., 2010), anal (Alemany et al., 2015), vaginal (Alemany et al., 2014), vulvar (de Sanjose et al., 2013), penile (Alemany et al., 2016) and HN cancers (Castellsague et al., 2016) are described elsewhere. For the purpose of this study, only squamous cell carcinoma samples from the vulva, penis and HN cancer that had tested negative for the presence of HPV DNA using the SPF10/DEIA protocol were included in the analyses. As positive controls, for each anatomical location we included further 5-10% samples of the same study that had tested positive for the presence DNA from a single HPV. The final dataset for the present work consisted of 1141 cases randomly chosen among all HPV DNA negative squamous head-and-neck cancers (380 cases from the larynx, 380 cases from the oral cavity and 381 cases from the pharynx) and 59 controls for squamous head-and-neck cancers, 902 cases and 83 controls for squamous vulvar cancers and 592 cases and 57 controls for squamous penile cancers.

*PCR and sequencing*

DNA was released from FFPE material by incubation of four 5µm slices with 250 µL of Proteinase K solution (10 mg/mL proteinase K, 50 mM Tris-HCl, pH 8.0) overnight at 56ºC. Samples were subsequently incubated at 95ºC for 8 minutes to inactivate proteinase K and stored at -80ºC until use. To serve as a control for the presence of input DNA, the human tubulin gene was targeted to generate an amplicon of 65 bp, the same length as the one generated by the SPF10 primers on the HPVs genomes (Alemany et al., 2015). For vulvar and penile cancer samples, the DNA solutions obtained after proteinase K treatment had been stored for several months at -80 ºC. For samples from these locations, to facilitate the release of DNA adsorbed to the plastic walls, tubes were heated at 60ºC during 48h prior to aliquot withdrawing for PCR.

Samples were analyzed using different sets of previously described primers, listed in Table S1, designed to detect a broad range of mucosal and cutaneous PVs: i) CPI/CPII (Tieben et al., 1993); ii) FAP6085/FAP64 (Li et al.); iii) MY9/MY11 (Manos et al., 1989); iv) SKF/SKR (Sasagawa and Mitsuishi, 2012). Additionally, we designed a new set of broad-HPV primers by using CODEHOP (Rose et al 2003), (MFI/II) specifically targeting the *E1* gene of cutaneous AlphaPVs. The MFI/II primer set was designed in order to complement the HPV detection of the SK primer set, as not all cutaneous AlphaPVs can be detected by using the SK primer set.

All PCRs reaction mixtures contained: 0.05 U/µL DNA Polymerase (Biotools), 1.0 x PCR reaction buffer (10X), 1.5 mM MgCl2, 0.4 mM dNTPs (Invitrogen), 0.2 µM of each primer (Biolegio) and 100 ng of DNA. Each PCR mixture was underwent 40 amplification cycles with different annealing temperatures for each primer set: 45ºC for SKF/R; 50ºC for CPI/II, FAP6084/64, MFI/II; and 47ºC for MY09/11. Finally, PCR products from those samples with an amplicon were sequenced at the Genoscreen facilities (Lille, France) in both strands, using the same primers used for amplification.

For all samples previously classified as SPF10/DEIA-negative in which we identified by Sanger-sequencing the presence of DNA from an HPV included in the 69 AlphaPVs detected by DEIA, we additionally performed a fresh SPF10 PCR followed by the LiPA25 genotyping assay (Kleter et al., 1999).

**RESULTS**

We have tested a total of 2,635 FFPE surgical samples from squamous carcinomas of the HN, penis and vulva that had previously tested negative for the presence of HPV DNA using the SPF10/DEIA procedure.

For squamous cancers of the HN, we were able to properly identify and genotype the HPV-DNA present in 31 out of 59 controls (53%) (Table 1). In all cases, the HPV hereby identified matched the one previously genotyped by LiPA25. We tested 1141 SPF10/DEIA-negative HN cancer samples, and we retrieved sequences specific for a one particular HPV in 15 out of these 1141 (1.3%) samples (Table 1, Table S2). In ten out of these 15 samples (66.7%), we detected DNA from mucosal HPVs, namely HPV16 (n=8), HPV51 (n=1) and HPV74 (n=1). Finally, we recovered DNA from two BetaHPVs, namely HPV20 and HPV21, and from three cutaneous AlphaPVs, namely HPV2, HPV57 and HPV61. Although ten from these 15 samples were expected to have tested positive for the initial SPF10/DEIA screening (i.e. those containing either HPV16, 51 or 74, targeted by the SPF10/DEIA) they resulted negative. When we performed a new SPF10/LiPA25 test, we recovered five of them (50%) as positive for HPV-DNA, matching in all cases the HPV genotype identified with the broad-spectrum primers, including HPV16 (n=3), HPV51 and HPV74.

For vulvar cancer samples, we were able to identify and genotype 62 out of 83 (75%) SPF10/DEIA-positive control samples. We tested further 902 SPF10/DEIA-negative samples and found HPV-DNA in 30 of them (3.3%) (Table 1, Table S3). One of the samples contained DNA of one cutaneous HPV, namely HPV2, while the remaining 29 samples contained DNA from essentially mucosal HPVs, namely HPV16 (n=18) HPV33 (n=1), HPV45 (n=2), HPV52 (n=1), HPV53 (n=1), HPV56 (n=1), HPV66 (n=1), HPV70 (n=1) and HPV74 (n=3). All these 29 samples should have tested positive for the initial SPF10/DEIA screening, and indeed 16 of them (55%) resulted positive when we performed a new SPF10/LiPA25 on them, the results matching those obtained with the broad primer sets. Specifically, we detected HPV16 (n=9), HPV33 (n=1), HPV45 (n=2), HPV52 (n=1), HPV70 (n=1) and HPV74 (n=2).

Finally, for penile cancer control samples, we successfully identified the presence of HPVs DNA in 37 out of 57 (65%). Among the 592 SPF10/DEIA-negative samples, we identified the presence of HPV DNA in 16 of them (2.7%), and in all cases the retrieved sequences corresponded to HPV16 (Table 1, Table S5). When we performed a new SPF10/LiPA25 genotyping on these 16 samples, four of them (25%) tested indeed positive and revealed the presence of HPV16 DNA.

**DISCUSSION**

We present here the largest study assessing the presence of DNA from cutaneous and mucosal HPVs in squamous cell carcinomas of the vulva, penis and HN, aiming at HPVs beyond those already identified as (probable/possible) oncogenic factors for these sites. We confirmed by multiple PCR tests the presence of DNA from a number of cutaneous HPVs in a very small number (6/2635) of cancer cases that seem to be truly negative for oncogenic HPVs. Nevertheless, it is important to keep in mind that the presence of viral DNA alone does not necessarily imply causation or relation to malignancy.

In the present study we have analyzed HPV-negative cancer cases of the HN, vulva and penis, previously tested by SPF10/DEIA/LiPA25 protocol. In HN samples, we detected HPV DNA in 15 out of 1141 (1.3%) samples previously diagnosed as HPV negative by SPF10/DEIA technology. Among them, five types corresponded to cutaneous HPVs: two belonging to BetaPVs (HPV20 and HPV21) and three belonging to AlphaPVs (HPV2, HPV57 and HPV61). All these samples came from the oral cavity except the HPV20, which was retrieved from a laryngeal cancer sample. In the same repository, we already confirmed the presence of two single infections by cutaneous HPVs (Table 3) (Castellsague et al., 2016). Studies on HPV prevalence demonstrated a high prevalence (more than 60%) of cutaneous HPVs in the oral cavity of healthy individuals (Bottalico et al., 2011; Lang Kuhs et al., 2013). Our results suggest that the contribution of these cutaneous types to cancer remains minor and may be neglected for screening and vaccination purposes. A prospective investigation (Agalliu et al., 2016) reported a positive association between infections with Beta and/or GammaPVs and increased head-and-neck cancer risk. Indeed, the authors reported that Beta1, Gamma11 and Gamma12 species were associated with 3.3 to 5.5-fold higher risk of HN carcinomas after adjusting for smoking, alcohol consumption and HPV16 detection. However, residual confounding cannot be ruled out as an explanation of these findings (Rollison and Gillison, 2016).

Other studies have also analyzed the presence of cutaneous HPVs in malignant head-and-neck tumours (Agalliu et al., 2016; Koskinen et al., 2003; Lindel et al., 2009; Paolini et al., 2012). Lindel and colleagues (Lindel et al., 2009) found a high prevalence of cutaneous HPVs in head-and-neck carcinomas samples (16/18 HPV-DNA positive tumours) by PCR amplification with multiple primers. The authors reported the presence of BetaPVs DNA as a single infection in eleven samples, and in two samples they found a multiple infection of two (or three) BetaPVs; finally, in three samples they found a multiple infection of a cutaneous BetaPV in combination with a low-risk

AlphaPV. Paolini and colleagues (Paolini et al., 2012) also demonstrated the presence of cutaneous BetaHPVs (HPV5, 14, 20, 21, 25, 36, 47, 100, 105 and 111) in HN squamous carcinomas samples (16/78 samples). However, they could not find cutaneous viral transcripts in any of these cancer samples.

We also analyzed penile SPF10/DEIA-negative cancer samples. Although in the analyzed samples, we did not find any cutaneous HPV type, the original description of the dataset already included the presence of cutaneous HPVs in two penile squamous carcinoma samples, namely HPV27 and HPV76 (Table3) (Alemany et al., 2016). Few published studies have examined the presence of cutaneous HPVs in penile carcinomas. Moreover, all have used varying methods and have focused on a single lesion type, making comparisons across studies difficult. By using PCR amplification with multiple primers, Wieland and colleagues (Wieland et al., 2000) described the presence of HPV8 (as single infection and as co-infection with a mucosal oncogenic type) in in situ squamous cell carcinomas of the penis (12/12 samples). In another study, the presence of HPV8 was found in penile squamous carcinoma samples, being that type the second most prevalent type (10/46 samples) as a single infection (6/10 samples) or as co-infection with a mucosal oncogenic type (4/10 samples) (Humbey et al., 2003). These authors also found other cutaneous BetaHPVs (HPV12, HPV17, HPV20 and HPV23) (Humbey et al., 2003).

Regarding vulvar squamous carcinomas, we detected the DNA of one cutaneous virus (HPV2) in one sample. De Koning and colleagues, by using PM-PCR reverse hybridization assay did not find any cutaneous HPV in a total of 39 vulvar cancer cases (de Koning et al., 2008). However, their study did not include squamous carcinoma samples in their set, making difficult the comparison of the results across the two studies.

Epidemiologic studies using standard methods of detection have demonstrated that other HPVs beyond high-risk HPVs can be involved in the malignization process (Guimera et al., 2013), although their prevalence is very low (Table 2). In the present study we describe the presence of HPV-DNA of cutaneous types in cancer samples. However, the mere presence of HPV-DNA in cancer samples is not sufficient to prove viral causation as it might simply reflect an ongoing viral infection unrelated to the carcinogenic process (Sarid and Gao, 2011). Thus, it is important to explore the expression patterns of other markers associated with HPV-induced carcinogenesis to assess the biological and oncogenic activities of HPVs identified in these cancers. Although the pathogenic pathway for malignant transformation in oncogenic AlphaPVs has been extensively studied (Bello et al., 2015; Duensing and Munger, 2004; Jones and Munger, 1997; Moody and Laimins, 2010; Munger et al., 1989), the putative oncogenic mechanism of cutaneous HPVs remains unknown and may probably involves different molecular pathways than those well described for HPV16 and other oncogenic HPVs (Groves and Coleman, 2015; Rusan et al., 2015).

In SPF10/DEIA-negative samples, we detected the DNA of one of the 69 types detected by the SPF10/DEIA technology (Kleter et al., 1999). Among the 1,141 HN SPF10/DEIA-negative samples, ten samples contained an HPV type detected by DEIA. In five out of these ten samples, the type was confirmed by performing a LiPA25 analysis. In vulvar cancer samples 29/902 samples resulted positive for types detected by DEIA, and the LiPA25 confirmed the HPV type detected by PCRs methods in 16 of them. Finally, for penile samples the LiPA25 detected the HPV type reported by PCR in four out of 16/592 samples. Our results suggest that the SPF10/DEIA/LiPA25 methodology present very small rates of false-negative cases (ranging between .0.9-3.5%), which confirm its high sensitivity, especially for FFPE samples (Geraets et al., 2015).

The fraction of SPF10/DEIA-positive samples detected by the PCR methods used in this study ranged from 53% in HN to 65-75% in penile and vulvar cancer samples, respectively, demonstrating that our results might be underestimating the presence of cutaneous HPVs, as well the number of SPF10/DEIA false negative samples. Globally, the broad-spectrum primers sets used in this study are not optimized for retrieving small amplicons. Primer selection has been done taking into account the spectrum of the HPV types detected, in order to cover both mucosal and cutaneous HPVs; and their sensitivity in a single round assay. Using this primer selection, we aimed to covering both mucosal and cutaneous HPVs, detecting both *E1* and *L1* genes, with an amplicon range from 150bp to 450bp. Nevertheless, the amplicon length was larger than the 65bp targeted by the SPF10 primers (Kleter et al., 1998). This difference in amplicon size can easily explain the incompletely recovery of positive SPF10/DEIA samples.

Another technical factor that might also contribute to a hypothetical underestimation of the presence of cutaneous HPVs is the storage of the samples at -80ºC, which might compromise the DNA quality of the samples (Lee et al 2010). In order to minimize this effect we heated all the stored samples (including samples from the vulva and penis), during 48h at 60ºC. We found that this pretreatment may help recover DNA possibly adsorbed onto the plastic tube walls, as the positivity of control samples were higher in the localizations were the pretreatment were performed. However, this recovery remained incomplete and still, samples with low DNA concentration may result in low DNA detection rates.

**CONCLUSIONS**

Our results show that previous estimates of HPVs involvement in vulvar, penile and HN cancers, based on SPF10/DEIA procedure do not suffer from significant underestimation biases. We show further that other HPVs beyond mucosal HPVs, typically detected by standard methods of detection may be present in carcinomas of the penis, vulva and HN, albeit their relative contribution remains minor and may have no public health impact whatsoever, and may be neglected for vaccination and cancer screening purposes. Nevertheless, our results brace previous reports with lower sample size, and mounting evidence suggests that certain cutaneous HPVs may be linked to a small number of cancer cases. Our results call for further studies to elucidate the pathogenic role of these "non-oncogenic" HPVs, the possible malignisation routes and mechanisms and the interplay between viral infection, host genetics and environment.

**FUNDING**

MFS is the recipient of an IDIBELL PhD scholarship. Partial support for this work was obtained from grants from the Instituto de Salud Carlos III (grants FIS PI081535, FIS PI1102096, FIS PI1102104, RCESP C03/09, RTICESP C03/10, RTIC RD06/0020/0095, RD12/0036/0056,), from the Agència de Gestió d'Ajuts Universitaris i de Recerca (2014SGR2016), from the Stichting Pathologie Ontwikkeling en Onderzoek (SPOO) foundation (the Netherlands), and from Sanofi Pasteur MSD and Merck & Co, Inc. None of the founders had any role in data analysis, interpretation, article writing or publishing.

**Aknowledgements**

We are grateful for the work of all the Institut Català d'Oncologia (ICO) team and the ICO International HPV in HN Cancer Study Group. The study is also part of the Human Papillomavirus Vulva, Vagina, Anus and Penis (HPV VVAP) international study coordinated at ICO, Barcelona, Spain.

**TABLES**

Table 1: Concordance between SPF10-DEIA and HPV broad-spectrum PCR (CPI/II, SKF/R, FAP6064/64, MY09/11 and MFI/II) in Head-and-neck (N=1200), Penile (N=649) and Vulvar (N=985) squamous cell carcinoma samples.

|  | Broad-HPV PCR | SPF10-DEIA | | | | Total |
|---|---|---|---|---|---|---|
|  |  | + | | - | | |
| **Head and Neck Squamous cell carcinoma** |  | N | % | N | % |  |
|  | + | 31 | 50.8 | 15[a] | 1.3 | **46** |
|  | - | 30 | 49.2 | 1126 | 98.7 | **1156** |
|  | Total | **61** | **100** | **1141** | **100** | **1202** |
| **Vulvar Squamous cell carcinoma** | Broad-HPV PCR | SPF10-DEIA | | | | Total |
|  |  | + | | - | | |
|  |  | N | % | N | % |  |
|  | + | 62 | 74.7 | 30[b] | 3.3 | **92** |
|  | - | 21 | 25.3 | 872 | 96.7 | **893** |
|  | Total | **83** | **100** | **902** | **100** | **985** |
| **Penile Squamous cell carcinoma** | Broad-HPV PCR | SPF10-DEIA | | | | Total |
|  |  | + | | - | | |
|  |  | N | % | N | % |  |
|  | + | 37 | 64.9 | 16[c] | 2.7 | **53** |
|  | - | 20 | 35.1 | 576 | 97.3 | **596** |
|  | Total | **57** | **100** | **592** | **100** | **649** |

[a] HPVs detected: HPV20 (n=1, BetaPV), HPV21 (n=1, BetaPV), HPV2 (n=1, AlphaPV), HPV57 (n=1, AlphaPV), HPV61 (n=1, AlphaPV), HPV16 (n=8, AlphaPV), HPV51 (n=1, AlphaPVs), HPV74 (n=1, AlphaPVs).
[b] HPVs detected: HPV2 (n=1, AlphaPVs), HPV16 (n=18, AlphaPVs) HPV33 (n=1, AlphaPVs), HPV45 (n=2, AlphaPVs), HPV52 (n=1, , AlphaPV), HPV53 (n=1, AlphaPV), HPV56 (n=1, AlphaPV), HPV66 (n=1, AlphaPV), HPV70 (n=1, AlphaPVs) and HPV74 (n=3, , AlphaPVs).
[c] HPVs detected: HPV16 (n=16, AlphaPV)

Table 2: HPV prevalence distribution found in anogenital and head and neck cancers. Information was obtained from the larger cross-sectional study on HPV prevalence distribution coordinated by the Catalan Institute of Oncology [9-14]. HPVs were classified according their phylogeny and clinical presentation: Mucosal oncogenic AlphaPVs, Mucocutaneous AlphaPVs causing genital warts, AlphaPVs causing cutaneous warts, Asymptomatic cutaneous BetaPVs and Asymptomatic GammaPVs.

| Genus | Clinical presentation | Percentage (%) | | | | | |
|---|---|---|---|---|---|---|---|
| | | Cervix (de Sanjosé et al. 2010) | Vulva (de Sanjosé et al. 2013) | Vagina (Alemany et al. 2013) | HN (Castellsague et al. 2016) | Penis (Alemany et al. 2016) | Anus (Alemany et al. 2015) |
| AlphaPVs | Mucosal oncogenic[a] | 98.39 | 96.61 | 91.57 | 94.28 | 89.16 | 94.25 |
| | Mucocutaneous, genital warts[b] | 1.09 | 3.39 | 4.42 | 5.30 | 8.73 | 2.65 |
| | Cutaneous warts[c] | 0.01 | 0.00 | 0.40 | 0.00 | 0.30 | 0.44 |
| BetaPVs | Asymptomatic, Cutaneous | 0.00 | 0.00 | 0.00 | 0.00 | 0.00 | 0.22 |
| GammaPVs | Asymptomatic, Cutaneous | 0.00 | 0.00 | 0.00 | 0.00 | 0.00 | 0.00 |
| Undetermined | | 0.50 | 0.00 | 3.61 | 0.42 | 1.81 | 2.65 |

[a]Mucosal oncogenic AlphaPVs: AlphaPV species 5, AlphaPVs species 6, AlphaPVs species7, AlphaPVs species 9 and AlphaPVs species 11
[b]Mucocutaneous genital warts AlphaPVs: AlphaPVs species 1, AlphaPVs species 8 and AlphaPVs species10
[c]Cutaneous warts AlphaPVs: AlphaPVs species 2, AlphaPVs species 3 and AlphaPVs species 4

Table 3: Number of single HPV infections found in head and neck, penile and vulvar carcinomas. For each anatomical location, the first column corresponds to the number of single HPV infections found by SPF10-DEIA-LIPA25 methodology [12-14]. The second column corresonds to the number of single HPV infections found in the negative SPF10/DEIA/LIPA25 samples from the corresponding repository in the same anatomical location.

| | | | Squamous carcinoma | | | | | |
|---|---|---|---|---|---|---|---|---|
| | | | Head and Neck | | Penis | | Vulva | |
| Clinical presentation | Specie | Type | Castellsagué et al. 2016 (original n=3680) | This study (retesting = 1141 previously negative) | Alemany et al. 2016 (original n=1010) | This study (retesting =592 previously negative) | de Sanjosé et al. 2013 (original n=1709) | This study (retesting =902 previously negative) |
| Mucosal lesions, oncogenic potential | Alpha-5 | HPV26 | 8 | 0 | 2 | 0 | 1 | 0 |
| | | HPV51 | 4 | 1 | 2 | 0 | 2 | 0 |
| | | HPV70 | 0 | 0 | 0 | 0 | 1 | 1 |
| | Alpha-6 | HPV30 | 1 | 0 | 2 | 0 | 1 | 0 |
| | | HPV53 | 2 | 0 | 2 | 0 | 3 | 1 |
| | | HPV56 | 1 | 0 | 2 | 0 | 6 | 1 |
| | | HPV66 | 1 | 0 | 2 | 0 | 1 | 1 |
| | Alpha-7 | HPV18 | 9 | 0 | 4 | 0 | 17 | 0 |
| | | HPV39 | 5 | 0 | 2 | 0 | 3 | 0 |
| | | HPV45 | 6 | 0 | 9 | 0 | 13 | 2 |
| | | HPV59 | 1 | 0 | 4 | 0 | 0 | 0 |
| | | HPV68 | 3 | 0 | 1 | 0 | 2 | 0 |
| | Alpha-9 | HPV16 | 339 | 8 | 194 | 16 | 318 | 18 |
| | | HPV31 | 2 | 0 | 2 | 0 | 5 | 0 |
| | | HPV33 | 11 | 0 | 2 | 0 | 25 | 1 |
| | | HPV35 | 11 | 0 | 9 | 0 | 1 | 0 |
| | | HPV52 | 5 | 0 | 4 | 0 | 8 | 1 |
| | | HPV58 | 4 | 0 | 3 | 0 | 3 | 0 |
| | | HPV67 | 2 | 0 | 0 | 0 | 0 | 0 |
| | Alpha-11 | HPV73 | 0 | 0 | 0 | 0 | 1 | 0 |
| | Total | | 415 (97.19%) | 9 (60%) | 246 (91.45%) | 16 (100%) | 411 (96.93%) | 26 (86.67%) |
| Usually benign, productive mucosal lesions | Alpha-1 | HPV32 | 0 | 0 | 2 | 0 | 0 | 0 |
| | | HPV42 | 0 | 0 | 1 | 0 | 0 | 0 |
| | Alpha-8 | HPV40 | 0 | 0 | 1 | 0 | 0 | 0 |
| | | HPV43 | 0 | 0 | 1 | 0 | 0 | 0 |
| | Alpha-10 | HPV6 | 5 | 0 | 9 | 0 | 5 | 0 |
| | | HPV11 | 2 | 0 | 4 | 0 | 2 | 0 |
| | | HPV13 | 3 | 0 | 0 | 0 | 0 | 0 |
| | | HPV74 | 0 | 1 | 2 | 0 | 3 | 3 |
| | Total | | 10 (2.34%) | 1 (6.67%) | 20 (7.43%) | 0 (0.00%) | 10 (2.36%) | 3 (10.00%) |
| Usually benign, productive cutaneous lesions | Alpha-3 | HPV61 | 0 | 1 | 0 | 0 | 2 | 0 |
| | | HPV102 | 0 | 0 | 0 | 0 | 1 | 0 |
| | Alpha-4 | HPV2 | 0 | 1 | 0 | 0 | 0 | 1 |
| | | HPV27 | 0 | 0 | 2 | 0 | 0 | 0 |
| | | HPV57 | 0 | 1 | 0 | 0 | 0 | 0 |
| | Alpha-14 | HPV90 | 1 | 0 | 0 | 0 | 0 | 0 |
| | Total | | 1 (0.23%) | 3 (20.00%) | 2 (0.74%) | 0 (0.00%) | 3 (0.71%) | 1 (3.33%) |
| Usually Asymptomatic Cutaneous | Beta-1 | HPV19 | 1 | 0 | 0 | 0 | 0 | 0 |
| | | HPV20 | 0 | 1 | 0 | 0 | 0 | 0 |
| | | HPV21 | 0 | 1 | 0 | 0 | 0 | 0 |
| | Beta-3 | HPV76 | 0 | 0 | 1 | 0 | 0 | 0 |
| | Total | | 1 (0.23%) | 2 (13.33%) | 1 (0.37%) | 0 (0.00%) | 0 (0.00%) | 0 (0.00%) |